# Measurement of radiation coherence by means of interference visibility in the reflected light

E. A. Tikhonov, A. K., Lyamets, Institute of physics of the National Academy of Sci., Kiev, Ukraine, e-mail: etikh@live.ru

**ABSTRACT** – Proposed, justified and tested measuring of beam spatial coherence, based on the detection of an interference visibility of the equal to intensity of beam replicas emerging under reflection from the rotated plane-parallel plate. The method consists in measuring the interference contrast at the simultaneous change of the longitudinal path difference and transverse shift of named replicas. Accurate measurement of the angle of rotated plate and the amplitude distribution of interference minimaxes allows to determine the width of the beam spatial coherence. Spatial coherence of some types of multimode lasers has been measured by the method.

## *1. Introduction*

Light interference in parallel beams uses for the solution of many scientific and practical problems. Lasers and interferometers represent a solid example of an embodiment of the interference in instrumentation. Laser interferometry, interference spectroscopy and holography are well-known scientific research directions and applications, based on the interference /1,2/.

One of the objectives of the light characterization is the determination of the coherence degree (spatial-temporal correlation). This task is also solved by creating, measuring and analysis of the interference visibility contrast (VC) .

As the basic device of temporary coherence measurement serves Michelson interferometer (MI). The indirect way of measurement of temporary coherence consists in measurement of radiation spectral width because the spectral distribution of radiation intensity and visibility of interference pattern are functions coupled by Fourier transformation.

Modifications of Young interferometer (YI) with beam interference from two spaced slits in crossed or parallel beams /3, 4, 5, 6/ are applied typically for measurement of space coherence. Changing distance between slits and registering visibility in the interference plane, it`s possible to determine degree of mutual correlation between spaced areas of wave front – the spatial coherence of radiation. The indirect method of definition of spatial coherence is based on measurement of angular divergence of radiation in compliance with van Cittert-Zernike theorem that proves the space coherence is Fourier image of space distribution of light source /7/.

Distinction of interferometry schemes, in addition to way of initial beam duplication, is manifested in management of their optical path difference: variables - for beams in MI and permanent, but with variable distance between beam replicas - in Young's interferometer. Emphasized distinction of interference schemes allows to carry out separate measurement of temporary and space coherence which are mutually connected ever.

The important feature of interference schemes consists in way of the initial beam duplication in replicas with equal phases and intensities. All known approaches are separated on two groups: amplitude dividing on initial beam or wave front dividing it. Equal intensity of equiphased interfering replicas promotes to highest visibility and accurate measurement of coherence function $\gamma(x,\tau)$. Visibility contrast (V) and coherent function of radiation $\gamma(x,\tau)$ are interconnected by the following expression /7/,

$$V(x,\tau) = (2\sqrt{I_1 I_2})\gamma(x,\tau)/(I_1 + I_2) \qquad (1)$$

In given work the 2-beam interference at light reflection from plane-parallel plate is studied. Application of this interference results for determination of light coherence was the objective. Actually it is development of the classical 2- beam interference in near field at precision readout of turning angle of plane-parallel plate and measurement of angular dependences of its VC in the higher interference order. As it will be shown further, the similar scheme with precision readout of angular position of minimaxes contains also possibility of spectral characterization of beam emission at the preset parameters of instrumental plate in the rotary interferometer (RIN) or on the contrary, measurements of optical parameters of such plates at known light beam parameters (similar devices on classical schemes of interference are manufactured by Bristol Instruments company, USA). When to compare functioning of RIN with MI and YI including the interferometer with holographic Bragg grating /6/, it should be recognized advantages of the last in setup simplicity and stability of operation even in the presence of difficult removable vibrations.

## 2. Analysis of interference registered with RIN

The optical scheme of the rotary interferometer is presented on fig.1. At reflection of paraxial 1-2 beam diameter ∅ from transparent plane parallel glass with thickness T and refraction index n forms two equal intensity antiphase copies of 1-2 input beam, spaced on δ in reasonable range of incidence angles form.

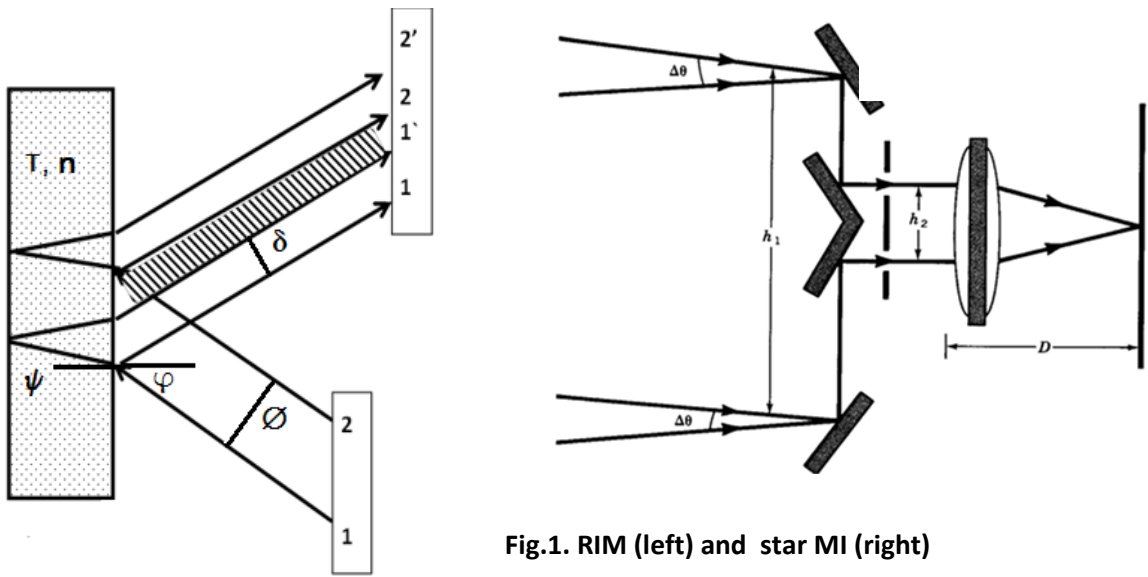

Fig.1. RIM (left) and star MI (right)

The beam 1-2 replica results at reflection on front edge of plane with $\pi$-phase jump, the 1'-2'-replica without phase jump is obliged to reflection on back edge of plate. The noted beams undergo to the longitudinal delay $\Delta$(mm) (or time delay $\Delta/c$) and transverse shift $\delta$ (mm) depending on incidence angle $\varphi$, refractive index n and thickness T of instrumental plate. Till with change of incidence angles the inequality remains $\delta<\varnothing$ in the region of overlapping of reflected beam couple interference exists and is recorded (the shaded zone in fig.1). In this configuration the photodetector registers interferential field with alternation of its power minimaxes as functions of incidence angle $\varphi$ at the set wavelengths $\lambda$ and optical parameters of instrumental plate.

Two-beam interference of parallel paraxial beams creates in near field diffraction rings with decreasing ring radius for the smaller orders if those are implemented at the given beam divergence. Therefore in order the measured VC responds to real values it is desirable to set the right aperture on the photo-receiver to cut the ring interference of lower orders. For typical laser beams the angular divergence is enough small and satisfies this condition without any filters. The registered picture of angular distribution of minimaxes remains also in far field diffraction when the beam focusing is applied before photodetector plane.

Raising of incidence angle results on beam replica overlapping decreases and disappears at all (in dependence on thickness T, index refraction "n" of plate and also beam diameter $\varnothing$ and beam divergence). The disappearance of interference in parallel beams with incidence angle $\varphi$ growth occurs at transverse shift value $\delta\geq\varnothing$ that is fixed by photodetector as disappearance of amplitude modulation of the reflected light power. It reminds disappearance of interference in star MI at increase a distance between of receiving mirrors more than value of light spatial coherence from the studied space object ($h_1$ on fig.1b.) /7/.

In configuration $\delta\geq\varnothing$ measurement of a coherence can be carried out from 2-dimensional interference pattern emerging in crossed beams with digital microscope /6/.

The nontrivial behavior of the interference visibility in area of decreasing of beam replica overlappings ($\delta\leq\varnothing$) on the way between the beam reflected plate and photodetector happens owing to addition of their mirror reversed parts of wave front. Really, for beams with limited spatial coherence across wave front the highest visibility is achieved at full presize mutual imposition of replica wave fronts, that takes place at the normal beam incidence /reflection only. Under increase transverse shift of the similar replicas their total antisymmetric overlapping and proportion of the in-phase wave fronts permanently decreases. It is clear that for partially coherent beam replicas cross shifts $\delta$ and connected time delay $\Delta$ connected with rotation of our plate (RIN) $\varphi$ will not influence on their interference visibility until $\delta$ and $\Delta$ stay small in compare to correspondent coherence parameters of the studied light beam. Moreover, for replica overlapping of the partially coherent beam without wave-front reversal, the increase the cross shift $\delta$ can follow even increase of VC due to growth of in-phase proportion in overlapping region.

Expressions for longitudinal wave path difference (delay) $\Delta$ and beam cross shift $\delta$ in the optical scheme of fig. 1a. can be written as :

$$\Delta = 2Tn\cos(\psi) = 2T\sqrt{n^2 - \sin^2(\varphi)}$$
$$\delta = 2Tn\cos(\varphi)\tan(\psi) = Tn\sin(2\varphi)/\sqrt{n^2 - \sin^2(\varphi)} \qquad (1a, b)$$

For understanding of ensuing discussion on fig. 2a, graphic dependences $\delta$ and $\Delta$ on incidence angle $\varphi$ on plates of different thickness of T=(2, 10, 15)mm at given IR=1,5 are given. It is seen that the magnitude and direction of change $\Delta$ and $\delta$ from $\varphi$ greatly different: change of cross shift $\delta$ in limits of incidence angles $\varphi$ from 0 to ≈0,8rad is much substantially than a changes of longitudinal path difference $\Delta$ in the same angular sector. This result gives the first indication to attribute the observed variations of interference visibility with $\varphi$ to spatial coherence. Really, absolute value $\delta(\varphi)$ at $\varphi=0\to0,8$rad rise multiply while $\Delta(\varphi)$ as responsible magnitude of time coherence in the same angular scan drops a little about 20% or less.

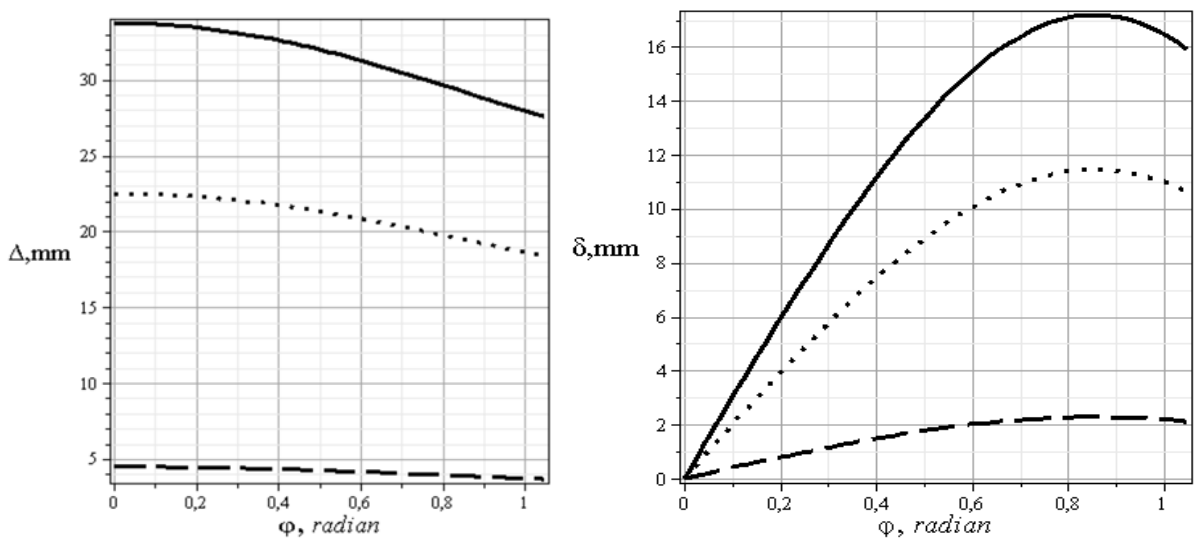

**Рис.2а,б. Изменение разности оптических путей $\Delta$ и поперечного сдвига $\delta$ в отраженном свете при повороте пластинки при T=2,10,15мм и n=1,5**

It is known that the optical path difference $\Delta$ defines the angular period of the interference pattern at the chosen plate parameters. Reduction of optical path difference with $\varphi$ in similar measurements means that influence of time coherence on VC decreases. However similar reduction of path difference does not eliminate the manifestation of space coherence. Only when $\Delta/c$ > time coherence magnitude of VC gets so small at initial angles, that quantitative measurement of spatial coherence is difficult to conduct. But when $\Delta/c$ < time coherence of studied radiation this new scheme can be compared with YI and it allowes identify the inherent transverse shift $\delta$ as adjustable distance between slots (beam replicas).

Thus, the considered characteristics of RIN give the some grounds to refer the registered changes of VC with incidence angle to influence of spatial coherence at almost switched off (more precisely - decreasing) manifestation of time coherence. However, if coherence length of the analyzed radiation is <<1 mm, measurement of space coherence requires the corresponding reduction of a longitudinal path difference for RIN determined by the set thickness T of the instrumental plate. It will lead to reduction of transverse displacement of the replicas $\delta_m$ with a turning angle $\varphi$ that can not allow to determine the size of space coherence area $\delta_c$ if the following inequality $\delta_c > \delta_m$ will run.

According to classical understanding of 2-beam interference of plane harmonic wave at wave path difference $\Delta(\varphi)$ on the set reflection angle $\varphi$ is described by formula (2) and it is represented graphically in fig. 3a:

$$I_\Sigma = 2I(1+\cos(2\pi\Delta(\varphi)/\lambda)) = 4I\cos^2(\pi\Delta(\varphi)/\lambda) \quad (2)$$

Dependence of the angular period of an interference on parameters of a plate, wavelength and the angle of registration can be used for measurement of plate thickness T or its index refraction. Accuracy of a such measurements is limited by the being available accuracy of the conjugated parameters: wavelengths and n or T. For receiving a convenient formula of processing the similar measurements, the dependence (2) allowing to find angular the period of interferential minimaxes as function of a turning angle of a plate of $\varphi$ is used:

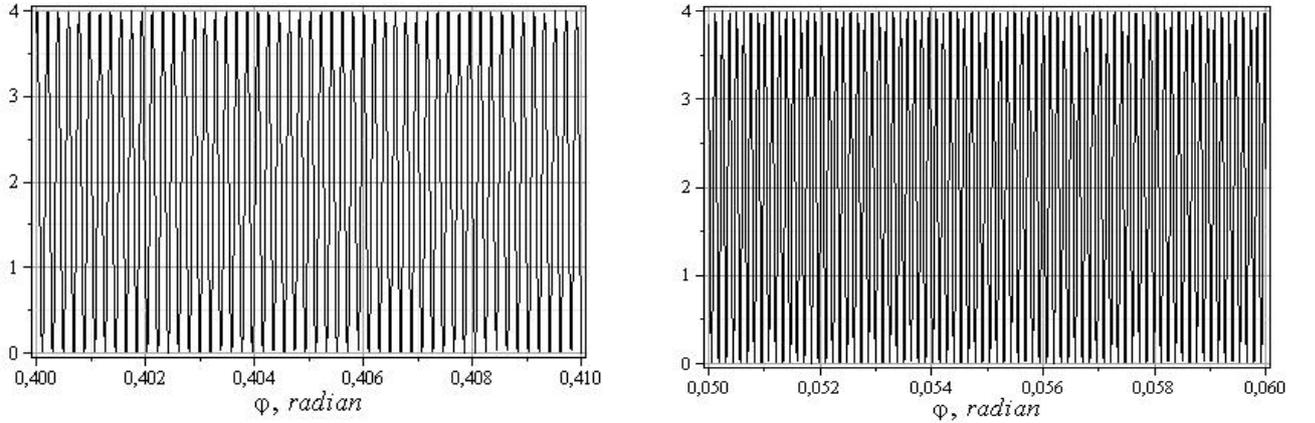

**Fig. 3. 2-beam interference in the reflected coherent light (λ=500nm) with plate T=2mm, n=1,5 for 2 different (difference 1 order) angles of scanning. Reduction of the angular period with φ at invariable 100% - visibility is observed.**

At $(\pi\Delta(\varphi)/\lambda) = (2m+1)\pi$ expression (2) has visibility maxima for m=0,1,2.. Then optical path difference for two next maxima is equal:

$$(2Tn/\lambda)(\cos(\psi_2) - \cos(\psi_1)) = 1 \quad (3)$$

Elementary transformations (3) allow to find the form (4) suitable for experimental check of the angular period of minimax alternation :

$$\varphi_2 - \varphi_1 \cong \lambda n / 4T\sin((\varphi_1 + \varphi_2)/2) \quad (4)$$

The formula (4) from measured data of the angular period $\varphi_2 - \varphi_1$ and their average of beam incidence angles on RIN (that corresponds to an average on the angles of emergence of the next maxima) allows to find any of the conjugated parameters: $\lambda$, T, n.

Further we will address the analysis of other factors which, in addition to coherence, manifest in VC at its registration by RIN. In fig. 4. the typical result of visibility record with RIN: angular distribution of power of the interferential field (minimaxes) for radiation of GaAlAs diode laser on $\lambda$=660nm is provided.

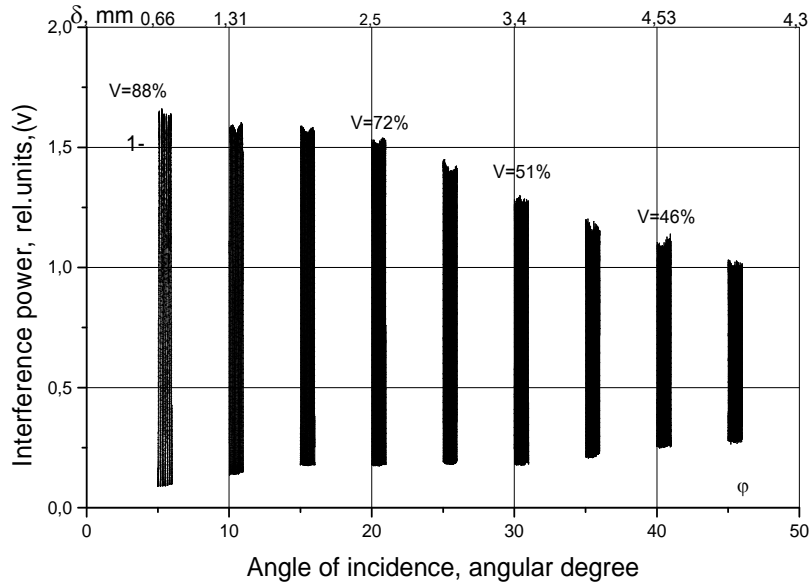

**Fig. 4.** Visibility of interference: radiation of diode GaAlAs laser with $\lambda$=660nm, parameters of RIN: T=1,9mm, n=1,52, ∅=3мм, $\alpha \leq 0,1'$- (wedge between planes of plate)

Envelope of minimaxes (over columns) show change of visibility contrast (VC) on the angle of scanning. It is evidently seen that VC extrapolation to normal angle of incident indicates its aspiration to VC$\cong$100%. For reduction of scanning time of area 0÷50$^0$ by step in some angular seconds continuous scanning was applied only in sectors $\approx$1÷2$^0$ a width. In these sectors of scanning harmonious modulation of power like to fig. 3.takes place. It is characterized by minimax alternation period of some minutes therefore in fig. 4. dense data recording at computer drawing forms "black boxes". Harmonious modulation becomes visible or at increase in angular resolution up to some milliradians as in fig.3. or at increase in the angular period of an interference on small angles 1÷2 degrees operation RIN (see a formula (4)).

CV is defined by the standard form as the relation of a difference of minimaxes to their sum. Its value in %-calculation for narrow sectors of scanning where it is almost constant and presented along columns on fig. 4. Level of the mean reflected power in replicas of the analyzed beam at change of an incidence angle in area to (0÷50)$^0$ (fig. 4.) remained thanks to the "successful" choice of an azimuthal angle for linear polarization of incident radiation, however at high linearity of a photo response of the registration scheme the value of VC did not depend on the level of total power.

Spatial dimension of cross shift $\delta$ on an incidence angle $\varphi$ has nonlinear character (formula (1a) and fig. 2a). Therefore, for an assessment of spatial coherence in spatial dimension the cross shift $\delta$ values on an upper scale of fig. 4. only for discrete values of an scanning angle φ are given. But till we will refrain from metrological estimates of space coherence and instead we will pay attention to other important feature of RIN functioning, in particular, to VC dependence from wedge angle plates used in the rotary interferometer RIN.

Striking reduction of VC interference for the radiation of the same diode GaAlAs laser with RIN on a wedge glass plate of T = 2,98mm and α =2' is presented in fig. 5. Dependences for VC in fig. 5. is recorded for two different azimuthal angles when weight deposits from S and P components of polarization in the reflected light change when incidence angle scans.

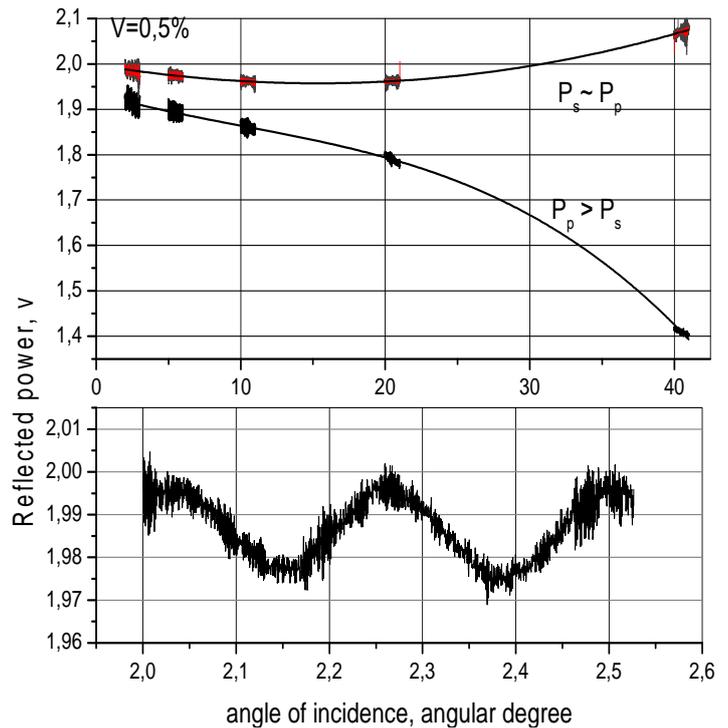

**Fig. 5. Decrease VC (φ) of the same diode laser emission at using a wedge plate with α=2'. In the lower drawing VC is unrolled with the greiter angular resolution of 6'.**

Total power of the lower curve decreases on the incidence angle because of a bigger contribution to the power brings p- polarized component. VC of interference with the wedge plate subsides to values less than 1% having come nearer to a noise level of the photo-registering scheme.

Similar reduction of VC is clear on the qualitative level. For RIN with a wedge plate it can be explained by lowering of an interference order of "m": $T\cos(\alpha)=m(\alpha)\lambda/2$. This lowering of interference order at a wedge equal 2' and the T =2,98mm makes 0,58% that is agreed with reduction of VC at comparison to VC for a plane-parallel plate (fig.4.). Therefore under appeal to the similar interferometry it is necessary to choose instrumental plates with wedge angles no more than 10" to realize high the right results of measurements.

Last metrological remark is about the role of spatial filtering of interferometric pattern which is recorded. Registration of VC by photodetector without spacial filtering also can reduces the actual contrast if the wave front set contains several overlapped interferential minimaxes at the same time. Results of fig. 6. show the photo-response of RIN in the similar case: application even of the slot diaphragm (not circular!) on the photodetector window increases VC in a nearfield of multi-mode radiation of HeNe, $\lambda$= 632,8nm appreciably.

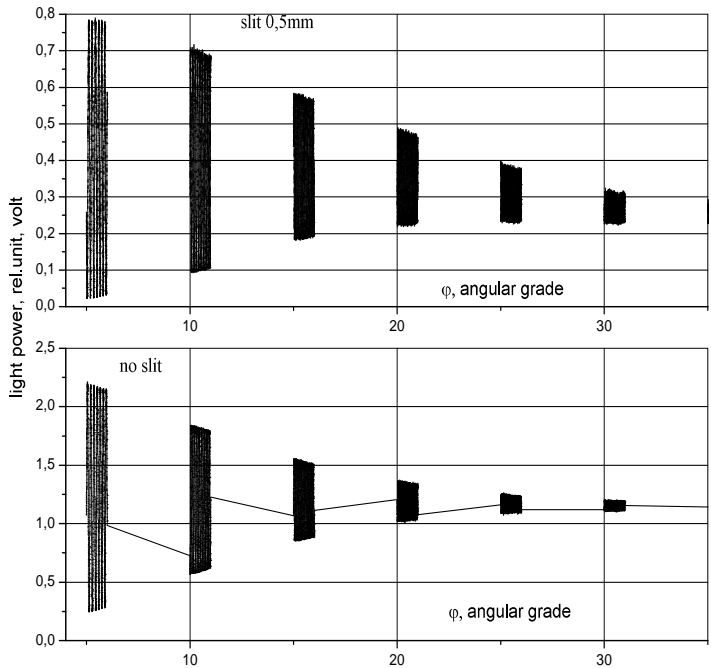

**Fig.6. Effect of a slot diaphragm before photoreceiver on VC of multimode HeNe laser: a) slot 0,5mm on a photodiode window: VC($5^0$) =99%, VC($15^0$)= 51% b)no slit: VC ($5^0$)=71%, VC ($15^0$)=45%**

However at formation of more difficult interference patterns than simple ring structure with decreasing of the angular period according to expression (4), positive influence of the slot diaphragm to magnitude VC of a such interference field ceases.

Further we will address comparative analysis of measurements of spatial coherence according to the scheme in fig. 1a. with the scheme of star Michelson interferometer (SMI), fig. 1b. In the SMI spatial coherence of radiation from a space objects is measured for determination their angular and linear sizes /7/. The value of spatial coherence of a radiation source $\varnothing_c$ is connected with the diffractional angular divergence by a ratio: $\Theta=\lambda/\varnothing_c$. Therefore the calculation of coherence based on spatial-temporal equivalence of optical path differences is possible. Owing to the equivalence functions of temporary and spatial coherence can be founded when to appeal to Fourier transformation of spectral or spatial intensity distribution $P(\omega)$, $P(\xi)$ respectively /7/ :

$$\gamma(\xi) = \int_0^\infty P(\xi)\cos\delta\, d\xi \qquad (5)$$

From (5) follows that for zero longitudinal difference of optical paths (at equal shoulders of an interferometer) when $\delta =0$, degree of space coherence of $\gamma(0)$=100%.
Interferometrical schemes of fig.1a,b. have similarities and distinctions: an interference in parallel paraxial beams with a changeable wave path difference $\Delta$ ($\varphi$) in our scheme and zero wave path difference - in SMI; the alternation period of interferental minimax in RIN is defined by an incidence angle $\varphi$ and other plate parameters; the same time VC depends on transverse shift value $\delta(\varphi)$ of the initial beam replicas. Transverse shift $\delta$ is equivalent to $h_1$-distance value between receiving mirrors of SMI for input beams on fig.1b: in SMI an interference disappears when distance $h_1$ between the receiving mirrors for two receiving beams exceeds cross size of spatial coherence of radiation in the place of SMI /7/.

The total wave path difference into SMI (fig. 1b) is the sum of the next two member: $\Sigma = (2\pi h_1 \Delta\Theta/\lambda + 2\pi h_2 x/\lambda D)$, where $\Delta\Theta = \lambda/h_1$ - angular divergence of radiation determined with space coherence, last member is seen from the given optical scheme of SMI. By analogy with SIM modelling we calculate a total wave path difference for our interferometer (fig.1a.):

$$\Sigma = (2\pi/\lambda)\delta\Delta\theta + (2\pi/\lambda)\Delta(\varphi) \qquad (6)$$

For the case in (6) member $2\pi h_2 x/\lambda D$ connected with specifical registration of an interference by SMI is absent, however appears member $\Delta$, connected with synchronous longitudinal shift that is resposible for descriptuion of time coherence. Substitution of a total optical wave path difference (6) in the main equation of a 2-beam interference (2) leads us to the following retio:

$$I_\Sigma = 2I(1+\cos\Sigma) = 2I(1+\cos(2\pi\delta/\otimes_c))\cos((2\pi/\lambda)\Delta(\varphi))$$
$$-\sin(2\pi\delta/\otimes_c))\sin((2\pi/\lambda)\Delta(\varphi)) \qquad (7)$$

Here the angle of diffraction of $\Delta\Theta=\lambda/\otimes_c$ is expressed through the required linear size of beam coherent area. At $\varphi=0$ expression (7) passes into in (2) due to zero cross shift with extreme VC=100%. The relation $(\delta/\otimes_c)$ can accept the integer values determining sizes multiple to the size of beam coherent area so at change of $\delta(\varphi)$ the corresponding extrema of distribution of $\gamma(\delta)$ can be registered. So, in point $\delta/\otimes_c=0,5$  $I_\Sigma=2I(1-\cos((2\pi/\lambda)\Delta(\varphi)))$, in point $\delta/\otimes=1$  $I_\Sigma=2I(1+\cos((2\pi/\lambda)\Delta(\varphi)))$.

In the field of small values $\delta/\otimes_c$ when the member with a sine can be neglected and the member a cosine to provide power series, the ratio (7) takes a such form:

$$I_\Sigma = 2I(1+\cos\Sigma) = 4I(1-0.25(\pi\delta(\varphi)/\otimes_c)^2)\cos((2\pi/\lambda)\Delta(\varphi)) \qquad (8)$$

The member in round brackets defines reduction of VC as increase of cross shift between the reflected copies of the studied beam. When $(\pi\delta(\varphi)/\otimes_t)^2=2$ VC falls half so this width of area can be connected with beam spatial coherence in a near field zone of diffraction:

$$\otimes_c=\pi\delta(\varphi)/1,41 \qquad (9)$$

The ratio (9) shows on a quantitative proportion the cross shift of $\delta$ measured by means of the rotary interferometer and area of the space coherence determined by the diffraction theory.

## 3. *Measurements of space coherence of different lasers*

The results of functioning of RIN presented above and their analysis give a basis for representation and interpretation of behavior of space coherence of several types of diode lasers. In fig. 7a, demonstrative results of VC for a emission of the diode GaN $\lambda$=405nm laser and power 50mW are given. In comparison with other studied sources the measured contrast of interference (VC) GaN-laser discovers the strong dependence from the wave path difference $\Delta(\varphi)$ even quite small cross shift $\delta(\varphi)$. The result is considered as direct confirmation of time coherence manifestation. Dependence of VC for instrumentel plates T =1,9mm and 0,71mm under scan on incidence angle φ defining the same time longitudinal optical way difference and their cross shift indicate to the following:

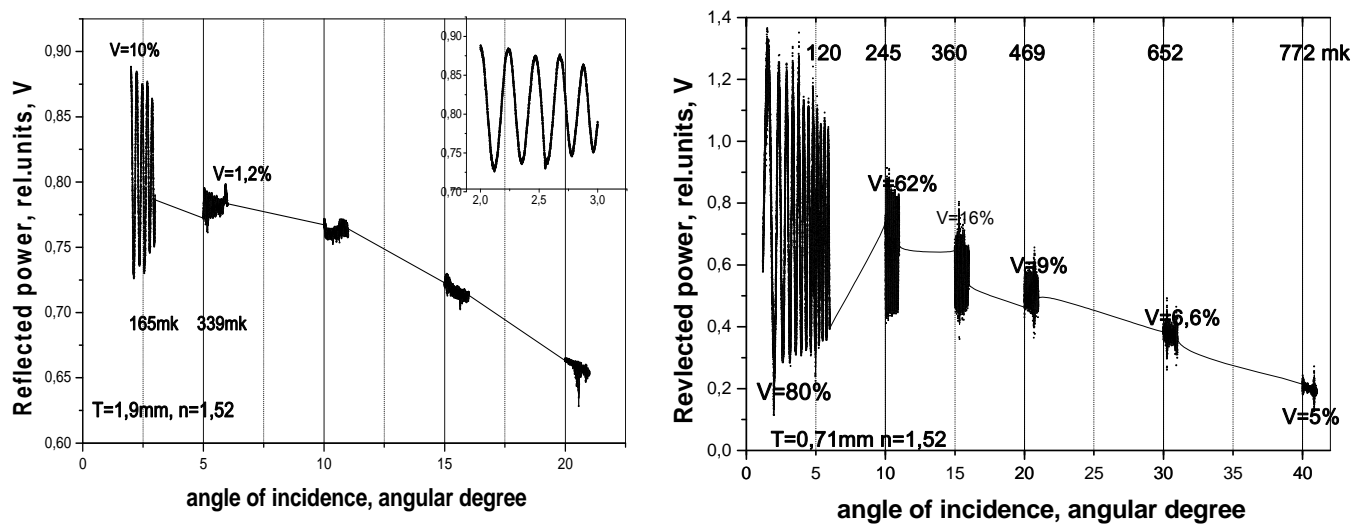

**Fig. 7a,b. Interference visibility of diode GaN laser ( 405nm) as function of a turning angle φ of RIN for 2 thickness of a measuring plate T =1,9mm (left) and 0,71mm (right)**

1) With the plate T =1,9mm even at small cross shift δ=165 mkm contrast of visibility falls to 10% and remaining invariable, contrary to expectations for the smaller cross shift. It is important to remind that increase in cross shift δ with an angle scan φ a longitudinal wave path difference reduces promoting manifestation of temporary coherence.

2) Reduction of thickness of measuring plate from 1,9mm to 0.71mm provides the sharp growth of VC on the same magnitude of replica cross shift with the plate T =1,9mm that obviously indicates commensurability of time coherence of radiation with a longitudinal delay of replicas with this instrumental plate.

3) From fig.7b. it is seen that in the sector of incidence angles from $2^0$ to $0^0$ value of VC strives to 100% that allows to exclude reduction of VC due to longitudinal delay on the bigger angles up to $40^0$ and to determine the spatial coherence width at the level of 50% VC=max approximately equal 250 microns.

4) Using measured VC data for $T_1$=0,71 and $T_2$ = 1,9mm at equal values of cross shift δ we find length interval in which to be coherence of this diode laser $2nT_2$> $L_c$> $2nT_2$ i.e. 2,16mm ÷5,78мм.

Similar determination of spatial coherence width for the GaAlAs diode laser with λ=660nm from data fig. 4. as area of reduce of VC from ≅100% on φ=0 to 50% on φ=$30^0$ results in much wider value δ=1550mkm!

For the correct comparisons of spatial coherence of different lasers as the region of reduce VC from 100% to 50% in micron main parameters of instrumental plane - thickness T, index refraction and wedge less 1' of instrumental plates are held invariable.
 (if the laser time coherence stays higher in compare the biggest time delay of RIM with given instrumental plate).

In fig. 8. VC measurements for other diode GaAlAs laser on λ=780 nm are presented. This GaAlAs laser is different from above studied at least other percentage ratio a component and, respectively, other lasing wave. The measurement shows 2-fold decrease of its spatial coherence width up to δ= 860mkm.

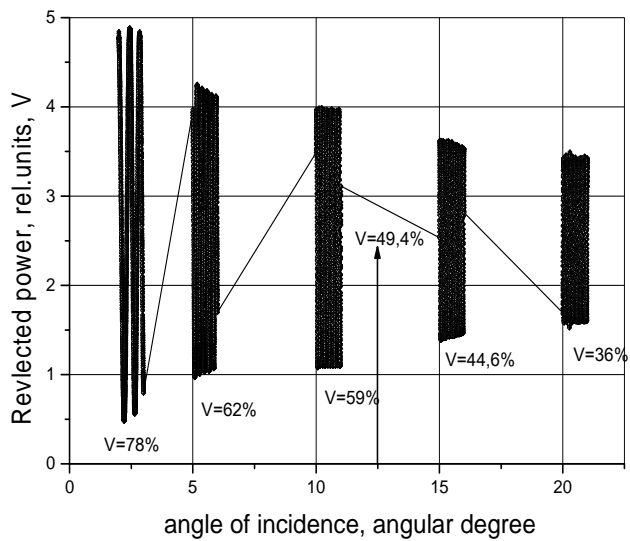

**Fig. 8. Contrast of visibility of diode GaAlAs laser 780 nm as function of cross shift of input beam replicas**

Similar estimates of the spatial coherence area were made for multimode HeNe (LGN-100) laser, λ=632,8nm. On the basis of VC data change above provided on fig.6a. spatial coherence width on the level of 50% reduce on the angle $15^0$ (with recalculation on linear cross shift) equals δ≅ 950mkm.

Above, in the second section of given work, all arguments on the basis of which observed change of VC for a number of lasers is referred to manifestation of space coherence were already declared. The most important of them is reduction of a wave path difference at turn of RIN towards big incidence angles (see fig. 2a.) from its maximum value on coal $\phi=0^0$. This reduction in percent relation to the greatest turning angles is defined by the minimum VC and makes 20-30% of an initial difference of the course. At the same time the replica cross shift δ of the analyzed beam connected to manifestation of its space coherence, increases from zero up to the maximum value on an incidence angle $\phi=45^0$, i.e. in infinite number of times (formula 1a cm, and fig. 2a, b). When measurements of GaN laser met with coherence length less than maximum wave path difference $\Delta_{max}=2Tn$, VC in that case became very low already on almost zero angles of measurement φ=δ~ 0. At comparable sizes of space and temporary coherence, VC of this radiation can remain invariable with a turning angle or increase if the space coherence is more than temporary. The received distinctions of the different lasers given on space coherence are not a subject of the physical analysis of our work as depend on a set of technical solutions in connection with their purpose.

## 4. Final notes

In this work the new technique of quantitative measurement of spatial/time coherence is offered, substantiated and approved. Approbation of a technique is carried out on typical diode and gas continuous-wave lasers. Possibilities of a technique in relation to pulse lasers are limited availability of the fast ADC devices for digitation of nano, - picosecond duration signals that is not principal restriction. As an necessary element of the similar measurement becomes a goniometer with precision reading and installation the angular turn and the synchronous angular rotation of reflected beams and photo-receivers of interference beam radiation.

Physical novelty of work at all simplicity of an experimental implementation its optical scheme consists in the appeal to an interference in parallel beams. It excludes need of the analysis of the complicated interference patterns localized in the crossing convergent beams used often at measurement of spatial coherence. Instead it VC of interference caused by variations of transverse shift of the imposed parallel beams with synchronous change of the longitudinal wave path difference connected with spatial and time coherence develops and records by PC like a progression of minimaxes. The choice of an optical thickness of a plane-parallel measuring plate it is possible to regulate in some limits a time delay between copies of the studied beams to avoid a confusing manifestation of time and spatial coherence simultaneously and realize the "pure" conditions for the measurement of single one of them.

Application of this method to laser beams allows to define both time coherence and, respectively, spectral width of radiation that connect us to Fourier spectroscopy and Fabry-Perot interferometry. Use of light sources with the calibrated values of wavelength allows to apply the described technique to measurement of some optical parameters of plane-parallel plates.

## *5. The references*